\documentclass[aps,pre,twocolumn]{revtex4-1}
\usepackage{amssymb}
\usepackage{amsmath,bm}
\usepackage{graphicx,color}
\newcommand{\B}[1]{{\bm{#1}}}%% Bold Roman & Greek Lower & Upper Case

\begin{document}

\title{Noise Amplification in Frictional Systems: Oscillatory Instabilities}

\author{Joyjit Chattoraj$^1$, Oleg Gendelman$^2$, Massimo Pica Ciamarra$^{1,3}$ and Itamar Procaccia$^4$ }
\affiliation{$^1$ School of Physical and Mathematical Sciences, Nanyang Technological University,
  Singapore\\$^2$ Faculty of Mechanical Engineering, Technion, Haifa 32000, Israel\\
  \normalsize{$^3$CNR--SPIN, Dipartimento di Scienze Fisiche,
    Universit\`a di Napoli Federico II, I-80126, Napoli, Italy}\\
  $^4$Department of Chemical Physics, the Weizmann Institute of Science, Rehovot 76100, Israel.}

\begin{abstract}
It was discovered recently that frictional granular materials can exhibit an important mechanism for instabilities, i.e the appearance of pairs of complex eigenvalues in their stability matrix.  The consequence is an oscillatory exponential growth of small perturbations which are tamed by dynamical nonlinearities. The amplification can be giant, many orders of magnitude, and it ends with a catastrophic system-spanning plastic event. Here we follow up on this discovery, explore the scaling laws characterizing the onset of the instability, the scenarios of the development of the instability with and without damping, and the nature of the eventual system spanning events.  The possible relevance to earthquake physics and to the transition from static to dynamic friction is discussed.
\end{abstract}

\maketitle

\section{Introduction}
Understanding how granular materials fail
is a problem of great geophysical importance due to its relevance to common phenomena
like earthquakes and landslides.
From a theoretical viewpoint, a micro-mechanical description of the failure process
is difficult due to the absence of a Hamiltonian description of granular systems.
To stress the importance of this, recall that
starting from the pioneering works of Malandro and Lacks \cite{98ML,99ML}
plastic failures in athermal amorphous systems (with forces derivable from a Hamiltonian) has been identified
as resulting from saddle node bifurcations occurring as a minimum of the energy
landscape becomes a saddle point as the system deforms.
In this approach, one considers a system of $N$ particles whose center of mass coordinates are $\B r_1, \B r_2, \cdots, \B r_N$ at temperature $T=0$ which is endowed with a Hamiltonian $U(\B r_1,\B r_2,\cdots\B r_N)$. The dynamics of the system is written as Newton's equations of motion:
\begin{equation}
m_{i}\frac{d^2 \B r_{i}}{dt^2}={\B F}_i(\B r_1,\B r_2,\cdots,\B r_N)\equiv -\frac{\partial U(\B r_1,\B r_2,\cdots
\B r_N)}{\partial \B r_i} \ .
\label{Newton1}
\end{equation}
When the system is in mechanical equilibrium the force $\B F_i$ on each particle vanishes,
 \begin{equation}
 \B F_i \equiv -\frac{\partial U(\B r_1,\B r_2,\cdots\B r_N)}{\partial\B r_i} = 0 \quad \text{in equilibrium} \ .
 \end{equation}
The stability against perturbations of the equilibrium state is determined by the second derivative of the Hamiltonian which is
the Hessian matrix:
\begin{equation}
H^{\alpha\beta}_{ij} \equiv \frac{\partial F^\alpha_i}{\partial r^\beta_j}\equiv-\frac{\partial^2 U(\B r_1, \B r_2,\cdots
\B r_N)}{\partial r_i^\alpha \partial r_j^\beta} \ .
\end{equation}
The Hessian matrix is evidently real and symmetric, and it has real eigenvalues which are all positive as long as the material is mechanically stable. Subjected to mechanical strain or stress the system may display a saddle node bifurcation with an eigenvalue going to zero. Generically this bifurcation results in the localization of the associated eigenfunction, in accordance with a local release of stress and energy \cite{04ML}. The modes of the Hessian matrix which are associated with such instabilities are known as ``plastic" or ``soft" modes and their probability density function differs from the usual Debye density of stats in purely elastic materials \cite{05Wya,10KLP}. Work was devoted to understand the system size dependence of the eigenvalues of the Hessian \cite{10KLP}, their role in determining the mechanical characteristics like the elastic moduli \cite{11HKLP}, the failure of nonlinear elasticity in such materials \cite{11HKLP,16PRSS,17DIPS}, and their relevance to shear banding and mechanical failure \cite{12DHP,13DHP,13DGMPS}.

The dynamics of granular systems is governed by interparticle interaction forces, which have both  normal and tangential components, obtained by coarse graining the microscopic degrees of freedom of the particles.
A variety of different models exist~\cite{49Min,Walton1986,JacquesMoreau1994,VuQuoc1999,VuQuoc2000,01SEGHLP},
reflecting the different assumptions made in the derivation of the effective forces.
In all cases, these forces cannot be expressed as derivatives of a Hamiltonian function with respect to the
coarse grained degrees of freedom, which are the translational and the angular displacements of the particles.
It is worth stressing that the lack of a Hamiltonian description is not exclusively related to the presence of viscous or hysteretic forces, but it is rather inherent in the coupling between the coarse grained tangential and normal forces.
This implies that granular materials lack a Hamiltonian description even when dissipative forces may play a negligible role. As a consequence, the micro-mechanics of the failure process of granular systems cannot be understood using Hessian methods. Hence the question arises: how do granular materials fail?

Previous works investigating the loss of mechanical rigidity of granular assemblies have
suggested that this could originate \cite{Nedderman1992}  from local events, related to the
discontinuity of the frictional interaction between macroscopic objects.
Indeed, the magnitude $|\B F_{ij}^{(t)}|$ of the tangential force acting between two macroscopic objects $i$ and $j$ is bounded by
$\mu |\B F_{ij}^{(n)}|$, where $\B F_{ij}^{(t)}$ is the normal interaction force and $\mu$ is Coulomb's friction coefficient.
Accordingly, contacts may reach their Coulomb threshold and start slipping as stress or stain is applied to a granular assembly. The failure of a contact could trigger subsequent rearrangements of other grains
in a avalanche process~\cite{Mcnamara2005,Welker2009}, or lead to the emergence
of a percolating cluster of unstable grains~\cite{CundallStrack1983, Staron2005, StaronRV2006},
hence induce the macroscopic failure of the system.
A complementary approach to rationalize the failure of granular systems has been
devised building on the theory for the instability of elastic-plastic solids~\cite{Hill1958},
and describes failure as a sharp increase of the kinetic energy of the system.
When a granular system is slowly deformed, work is done on the system.
Failure might occur when the system becomes unstable, meaning that the work done
on the system, and possibly other work done by the system, is converted into kinetic energy.
This so-called second-order work failure criterion~\cite{Nicot2014} is formalized at the level
of continuum equations, and it therefore not suitable to investigated the underlying microscopic
features leading to the instability. Nevertheless, this criterion is indirectly related to
the instability criterion developed for Hamiltonian systems, and hints towards the existence
of collective failure modes.
Support for the existence of a collective failure mechanism not initiated by the Coulomb failure of a contact
also comes from
detailed numerical simulations of model frictional granular particles, subject to an
increasing in shear stress.
These simulations have shown that a granular system may become unstable without
any new contact reaching its Coulomb threshold~\cite{Ciamarra2010}:
only after the system becomes unstable and starts slipping new contacts are seen to reach their Coulomb limit.
An indirect support towards the existence of collective failure mechanisms also stems from
the existence of a linear regime in the response of granular systems to applied perturbations
~\cite{96Mel,06FB,Griffa2013,Giacco2015e,Giacco2018}, which is not easily rationalized assuming failure
to originate from an underlying discontinuous process such as the Coulomb failure of a contact.

In a recent Letter~\cite{19CGPP} we identified an instability mechanism in systems lacking a Hamiltonian description, such as frictional amorphous solids. This instability mechanisms involves an oscillating exponential growth of deviations
from the state of mechanical equilibrium, and it is not related to the existence of a Coulomb threshold. It cannot exist in amorphous solids in which the inter-particle forces are derived from a Hamiltonian.
These oscillatory instabilities furnish a micro-mechanical mechanism for a giant
amplification of small perturbations that can lead to system spanning plastic events and mechanical failure.
This physics was demonstrated in the context of amorphous assemblies of frictional disks, but the mechanism is generic for systems with friction. In this paper we follow up and discuss the phenomenon in greater detail.

The structure of the paper is as follows:
in Section~\ref{cgmodel} we explicitly show that the coarse-grained description of the interparticle interaction of granular systems is not Hamiltonian, and introduce a modified interaction model that, while not-Hamiltonian, is at least differentiable.
In Section~\ref{oscillatory} we review the novel instability mechanism for frictional systems~\cite{19CGPP},
and then describe its critical features. Using the numerical model illustrated in
Section~\ref{model}, we test our theoretical predictions concerning the birth of the instability
in Sec.~\ref{linear}. We discuss the growth of the instability until nonlinear dynamics set in, culminating in a system spanning catastrophic event, in Sec.~\ref{develop}.
Section~\ref{damp} provides information on the effect of damping. The upshot of the discussion is that
a threshold damping frequency can be defined (based on the instability amplification frequency) below which the instability dynamics is unperturbed by damping. Sect.~\ref{howgen} raises up the important issue how generic is the instability
discussed in this paper. Does its existence depend on the details of the coarse grained model used, or do we expect
to appear generically in any models that employs a reduced set of coordinates like the positions of the centers of mass
of the granules and their angular coordinates. We provide arguments for the generality of the phenomenon but
propose that at this point in time experiments should be invoked as the final test.
In Sec.~\ref{summary} we summarize the paper and discuss the possible connection
of the present findings to remote triggering in earth quakes and to the transition from static to dynamic friction. The road ahead and future research are described.

\section{non Hamiltonian description of granular systems\label{cgmodel}}
Coarse grained descriptions of frictional amorphous solids, in which the effective degrees of freedom
are the translational and the angular positions of the particles, do not admit a Hamiltonian description.
To explicitly show that this is the case we recap here a popular and time honoured model used to describe the interaction between frictional granular particles, but the conclusion remains valid for other models.

The interaction between two particles, that we assume spherical for simplicity,
has a normal and a tangential component.
The normal force is determined by the overlap $\delta_{ij} \equiv \sigma_i+\sigma_j-r_{ij}$ between the particles,
where $\B r_{ij}\equiv \B r_i-\B r_j$ and $\sigma_i$ is the radius of particle $i$, and it is given by the Hertzian model,
\begin{equation}
\B F_{ij}^{(n)} = k_n \delta_{ij}^{3/2}\hat r_{ij} \ , \quad \hat r_{ij} \equiv \B r_{ij}/r_{ij}.
\label{Fn}
\end{equation}
%The reader should note that this force is
%a model force, and although its use is time honored, it should be considered as an effective
%force since we neglect the precise deformation of the contact area between the disks.
The tangential force is a function of the tangential displacement $\B t_{ij}$ between the particles, a vector which is always orthogonal to $\hat{r}_{ij}$. Upon first contact between the particles, $t_{ij}=0$. Providing every particle with the angular coordinate $\B\theta_i$ the change in tangential displacement is given by
\begin{equation}
d\B t_{ij} =d\B r_{ij} -(d\B r_{ij}\cdot \B r_{ij}) \hat r_{ij} +\hat r_{ij} \times (\sigma_i d\B\theta_i +\sigma_jd\B\theta_j) \ .
\end{equation}
Accordingly, $\B t_{ij}$ is obtained by integrating over time the relative velocity of the particles at the point of contact.
In the Mindlin model, the tangential force depends on $\B t_{ij}$ and on $\B \delta_{ij}$  \cite{49Min}
\begin{equation}
\B F_{ij}^{(t)} = -k_t\delta_{ij}^{1/2}t_{ij} \hat t_{ij} \ ,
\label{Min}
\end{equation}
and satisfy the Coulomb condition
\begin{equation}
\B F_{ij}^{(t)} \le \mu \B F_{ij}^{(n)} \ ,
\label{Coul}
\end{equation}
where $\mu$ is the friction coefficient. Usually this law is interpreted such that the tangential force reaches the limit abruptly, not analytically, thus not allowing to compute derivatives of the tangential force.

We can see now why the interaction is not Hamiltonian and the Hessian does not exist in this case. The first reason is somewhat trivial, stemming from the non-analyticity of the Coulomb law. This can be easily taken care of by smoothing out the Coulomb law such that the tangential force will have smooth derivatives; we choose:
\begin{eqnarray}
&&\B F_{ij}^{(t)} = -k_t\delta_{ij}^{1/2}\left[1+\frac{t_{ij}}{t^*_{ij}} -\left(\frac{t_{ij}}{t^*_{ij}}\right)^2\right]t_{ij} \hat t_{ij} \ , \nonumber\\
&&t^*_{ij} \equiv \mu \frac{k_n}{k_t} \delta_{ij} \ .
\label{Ft}
\end{eqnarray}
Now the derivative of the force with respect to $t_{ij}$ vanishes smoothly at $t_{ij}=t^*_{ij}$ and the Coulomb law Eq.~(\ref{Coul}) is fulfilled.

The second reason for the the loss of the Hessian matrix is not trivial at all \cite{Henkes2010}.
The time honored Hertz-Mindlin effective forces presented here, even for disks or balls, are not derivable from a potential, due to the coupling between the normal and the tangential displacement in the tangential force. Notice that this coupling is physical,
as the tangential force depends on the normal force since it determines the contact area. This is the origin of the
term $\delta_{ij}^{1/2}$ in Eqs.~(\ref{Min}) and (\ref{Ft}). It is easy to see that, because of this term,
the derivative of the normal force with respect to $t$ does not equal the derivative of the tangential force with respect to
$\delta$, which is what occurs when the forces are derived from a Hamiltonian $U(\delta,t)$.

\section{Oscillatory instabilities: theory\label{oscillatory}}
\subsection{Stability matrix}
While granular materials are not Hamiltonian, their dynamics is still Newtonian, with an
extended set of coordinates $\B q_i=\{\B r_i, \B \theta_i\}$:
\begin{eqnarray}
   \label{Newton2a}
   m_{i}\frac{d^2 \B r_{i}}{dt^2}&=&{\B F}_i(\B q_1,\B q_2,\cdots,\B q_N)\ , \\
   \label{Newton2b}
   I_{i}\frac{d^2 \B \theta_{i}}{dt^2}&=&{\B T}_i(\B q_1,\B q_2,\cdots,\B q_N)\ ,
\end{eqnarray}
where $m_i$ are masses for the center of mass coordinates, $I_i$ are moments of inertia for the angles, ${\B F}_i$ are forces and ${\B T}_i$ are torques, respectively.
Using the smoothed out force Eq.~(\ref{Ft}), this allows to define the stability matrix which is an operator obtained from the derivatives of the force $\B F_i$ and the torque $\B T_i$ on each particle with respect to the coordinates. In other words
\begin{equation}
J_{ij}^{\alpha\xi} \equiv \frac{\partial \tilde F^\alpha_i}{\partial q_j^\xi}\ , \quad \tilde{ \B F}_i \equiv \sum_j \tilde{\B F}_{ij} \ ,
\end{equation}
where $\B q_j$ stands for either a spatial position or a tangential coordinate, and $\tilde {\B F}_i$ stands for either a force or a torque. Since in the
usual case $\B F_i =-\partial U/\partial \B r_i$ we see that the operator $\B J$ is an analog
of the Hessian even when a Hamiltonian description is lacking.
But with a huge difference: $\B J$ is not a symmetric operator.
Being real it can possess pairs of complex eigenvalues.
When these appear, the system will exhibit
oscillatory instabilities, since one of each complex pair will cause an oscillatory exponential divergence of any perturbation, and the other an oscillatory exponential decay.
The actual calculation of the operator $J$ is somewhat cumbersome if conceptually straightforward. A detailed calculation
for the present case of frictional disks interacting via Eqs.~(\ref{Fn}) and (\ref{Ft}) is presented in Appendix~\ref{Jacobian}.

\subsection{The oscillatory instability}
When a pair of complex eigenvalues $\lambda_{1,2}=\lambda_r\pm i\lambda_i$ gets born, a novel instability mechanism develops. It should be stressed that the birth of a pair of complex eigenvalues is {\em not} a Hopf bifurcation. A pair of complex conjugate eigenvalues correspond to FOUR solutions $e^{i\omega t}$ to the linearized equation
of motion with
\begin{equation}
i\omega_{1,2} = \omega_i \pm i\omega_r\ , \quad i\omega_{3,4}=-\omega_i \pm i\omega_r\ ,
\label{four}
\end{equation}
with $\omega_r \pm i\omega_i = \sqrt{\lambda_r \pm i\lambda_i}$.
The first pair in Eq.~(\ref{four}) will induce an oscillatory motion with an exponential growth of any deviation $\B q(0)$ from a state of mechanical equilibrium,
\begin{equation}
 \B q(t) = \B q(0) e^{\omega_i t} \sin(\omega_r t).
 \label{growth}
\end{equation}
The second pair represents an exponentially decaying oscillatory solution. The actual spatial dynamics that sets in due to this instability will be discussed below in~\ref{secspiral}.

\subsection{Universal Scaling Laws~\label{sec:scaling}}
Noticing that the instability under discussion is somewhat unusual, we address the question how the coalescence
of two real eigenvalues and the bifurcation of two imaginary parts take place as a function of the imposed strain.
As noticed above, we need at least four degrees of freedom to have this instability, which means either four first order
equations or two second order equations. Since we are solving Newton's equations of motion we will discuss the instability
in the latter form. Since the bifurcation
is of co-dimension 1, it is enough to take into account a single parameter which we denote as $\gamma$ for obvious reasons. So, we consider the pair of equations:
\begin{equation} \label{eq1}
\begin{array}{l} {\ddot{x}_{1} +f_{1} (x_{1} ,x_{2} ,\gamma )=0} \\ {\ddot{x}_{2} +f_{2} (x_{1} ,x_{2} ,\gamma )=0} \end{array}
\end{equation}
We assume that both functions $f_{1} ,{\rm \; }f_{2} $ are analytic with respect to all their arguments.  Let us also assume that the system \eqref{eq1} has a family of equilibrium points $(\tilde{x}_{1} (\gamma ),\tilde{x}_{2} (\gamma ))$ for some interval of $\gamma $, so that
\begin{eqnarray} \label{eq2}
&&f_{1} (\tilde{x}_{1} (\gamma ),\tilde{x}_{2} (\gamma),\gamma )=0 \ , \\
&&f_{2} (\tilde{x}_{1} (\gamma ),\tilde{x}_{2} (\gamma ),\gamma )=0,\ \quad \gamma \in (\gamma_{1} ,\gamma_{2} ) \nonumber \ .
\end{eqnarray}
Next define the new variables $y_{1} =x_{1} -\tilde{x}_{1} (\gamma ),y_{2} =x_{2} -\tilde{x}_{2} (\gamma  )$. Then, the system \eqref{eq1} is rewritten as:
\begin{eqnarray} \label{eq3}
&&{\ddot{y}_{1} +F_{1} (y_{1} ,y_{2} ,\gamma  )=0} \\ &&{\ddot{y}_{2} +F_{2} (y_{1} ,y_{2} ,\gamma  )=0} \nonumber\\&& {F_{k} =f_{k} (y_{1} +\tilde{x}_{1} (\gamma  ),y_{2} +\tilde{x}_{2} (\gamma  ),\gamma  ) \ , k=1,2}\ . \nonumber
\end{eqnarray}
By construction Eqs.~\eqref{eq3} possesses equilibria $(y_{1} ,y_{2} )=(0,0)$ for $\gamma  \in (\gamma _{1} ,\gamma _{2} )$. Therefore, using analyticity, we rewrite equations \eqref{eq3} as follows:

\begin{equation} \label{eq4}
\begin{array}{l} {\ddot{y}_{1} +c_{11} (\gamma  )y_{1} +c_{12} (\gamma  )y_{2} +O(\left|y_{1} \right|^{2} ,\left|y_{2} \right|^{2} ,\left|y_{1} y_{2} \right|)=0} \\ {\ddot{y}_{2} +c_{21} (\gamma  )y_{1} +c_{22} (\gamma  )y_{2} +O(\left|y_{1} \right|^{2} ,\left|y_{2} \right|^{2} ,\left|y_{1} y_{2} \right|)=0} \\ {c_{kl} (\gamma )=\left. \frac{\partial F_{k} (y_{1} ,y_{2} ,\gamma  )}{\partial y_{l} } \right|_{(y_{1} ,y_{2} )=(0,0)} } \end{array}
\end{equation}
Removing the nonlinearities, and taking $y_{1} =y_1^{(0)} \exp (i\omega t),y_{2} =y_2^{(0)} \exp (i\omega t)$, one obtains the following equation for the eigenvalues:
\begin{equation} \label{eq5}
(c_{11} (\gamma )-\lambda )(c_{22} (\gamma )-\lambda )-c_{12} (\gamma )c_{21} (\mu )=0,{\rm \; }\lambda {\rm =}\omega ^{2} \ .
\end{equation}
The solution is obvious:
\begin{eqnarray} \label{eq6}
&&\lambda _{1,2} =\frac{c_{11} (\gamma )+c_{22} (\gamma )\pm \sqrt{D(\gamma )} }{2} \ , \\
&&D(\gamma )=(c_{11} (\gamma )-c_{22} (\gamma ))^{2} +4c_{12} (\gamma )c_{21} (\gamma ) \ . \nonumber
\end{eqnarray}

Finally we identify the critical point where the real eigenvalues coincide as $\gamma =\gamma_c ,\gamma_c \in (\gamma _{1} ,\gamma _{2} )$. Then, it must be $D(\gamma_c )=0$. In the vicinity of $\gamma_c $ we assume generic dependence of all functions on the parameters, and get the following:
\begin{eqnarray} \label{eq7}
&&\lambda _{1,2} =\frac{c_{11} (\gamma_c )+c_{22} (\gamma_c )\pm \sqrt{\alpha (\gamma -\gamma_c )} }{2} +O(\left|\gamma -\gamma_c  \right|)\ , \nonumber\\
&&\alpha = D'(\gamma)\Big |_{\gamma =\gamma_c} \ .
\end{eqnarray}

Without loss of generality, we can assume $\alpha >0$. Finally, for the eigenvalues near the bifurcation point we have:

\begin{eqnarray} \label{eq8}
&&\lambda _{1} -\lambda _{2} =\sqrt{\alpha (\gamma -\gamma_c )} +O(\left|\gamma -\gamma_c \right|)\ , \quad \gamma  > \gamma_c \\
&& \Im {(\lambda_1-\lambda_2)}= \sqrt{\alpha |\gamma -\gamma_c |} + O(\left|\gamma -\gamma_c \right|)\ ,\quad \gamma < \gamma_c \ . \nonumber
\end{eqnarray}

We thus conclude that the bifurcation is characterized by a square-root singularity for both the coalescence of the real eigenvalues and for the bifurcation of the imaginary parts. We verify this prediction in Sec.\ref{singularity}.
% The next subsection puts this prediction to the numerical test.

\section{Numerical model\label{model}}
To demonstrate the predictions presented above (which are based on a linearized stability matrix) we employ the following
numerical simulations. In addition, we will use numerical simulations to investigate the evolution
of the instability away from the linear regime, where no theoretical predictions are available, as well as the effect of damping.

We focus on a specific example of a binary assembly of $N$ frictional disks half of which with radius $\sigma_1=0.5$ and the other half with $\sigma_2=0.7$, unit mass $m=1$ and moment of inertia $I_i = 0.5 m_i\sigma_i^2$.
The normal interaction between the grains is given by Eq.~(\ref{Fn}), while
the tangential one is given by Eq.~(\ref{Ft}), with $k_t=2k_n/7$.
We use $m$, $2\sigma_1$ and $\sqrt{m(2\sigma_1)^{-1/2}k_n^{-1}}$ as our units
of mass, length and time, respectively. We consider different values for the friction coefficient
$\mu$ and the system size $N$.

The equation of motion are solved using two types of algorithms: ``Newtonian'' and ``Overdamped''. The first is simply a solution of the Newton equations of motion with the given forces Eqs.~(\ref{Fn}) and (\ref{Ft}).
The second algorithm is solving the same equations of motion but with a damping force that is proportional to the velocities of the disks with a coefficient of proportionality $\eta_v=m\eta_0$.
If not otherwise mentioned we use $\eta_0 = 2.2\times 10^{-2}$ expressed in reduced units.
This value of $\eta_0$ ensures that the dynamics is overdamped as the damping timescale $\eta_v^{-1}$ is of the order of the time that sounds needs to travel one particle diameter.
We use LAMMPS~\cite{Plimpton1995} to perform the numerical integration for these two algorithms,
with integration timestep $10^{-5}\sqrt{(2\sigma_1)^{1/2}k_nm^{-1}}$.

%%%%%%%%%%%%%%%%%%%%%%%%%%%%%%%%%%%%%%%%%%%%%%%%%%%%%
\section{Critical feature of the instability\label{linear}}
\subsection{Birth of complex conjugate pairs}
\begin{figure}
%\vskip 0.4 cm
    \includegraphics[width=0.35\textwidth]{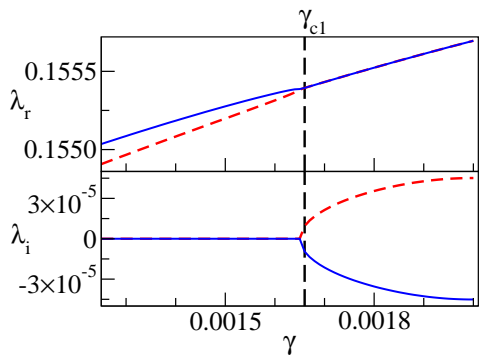}
    \caption{Upon increasing the strain $\gamma$ two modes with real eigenvalues $\lambda$ coalesce at $\gamma_{c1}$ (dashed vertical lines), and a pair of complex conjugate modes gets born. The upper and the lower panels show the evolution of the real and of the imaginary components of these modes for a system with $N=500$, $\mu=0.5$.}
    \label{bifurcation}
\end{figure}

To demonstrate the novel instability mechanism and describe its critical features,
we start recapping the results
of Ref.~\cite{19CGPP} with regard to the emergence of a pair of complex eigenvalues.
An initial configuration is prepared by arranging binary particles randomly in a two dimensional box and then perform two consecutive runs of overdamped dynamics to bring the configuration at mechanical equilibrium.
The initial configuration is prepared focussing on a frictionless system (i.e. $\mu=0$), and hence has no complex eigenvalues.
Afterwards, we switch on friction, and perform athermal quasi static (AQS) simulations: starting from the initial stable configuration we shear the simulation box along the horizontal direction ($x$) by the amount $\delta\gamma$ and then we run the overdamped dynamics until the system reaches mechanical equilibrium. Operatively, we consider the system to be in mechanical equilibrium when the net force on each particle is less than $5\times10^{-14}$.
%Due to shear each particle experiences an affine shift along $x$ based on their vertical coordinates $r^y_i$, i.e.~$\delta r_i^x = \delta\gamma r^y_i$.
Here $\delta\gamma$ varies in the range $10^{-4}$ to $10^{-6}$ depending on the precision needed for the identification of the instability.
After every AQS step we diagonalize the matrix $\B J$ to find its eigenvalues.
At some value of $\gamma$ we find for the first time the birth of conjugate pair of complex eigenvalues as seen in Fig.~\ref{bifurcation}.
%%%%%%%%%%%%%%%%%%%%%%%%%%%%%%%%%%%%%%%%%%%
If we continue to increase the strain using the same protocol, we see the continuous emergence of other complex pairs, as well as the death of existing ones. We remark that we observe complex eigenvalues, which correspond to an unstable system, as we are using in this investigation an overdamped dynamics, which kills the growth of the instability. The effect of damping is discussed in Sec. \ref{damp}

%%%%%%%%%%%%%%%%%%%%%%%%%%%%%%%%%%%%%

%%%%%%%%%%%%%%%%%%%%%%%%%%%%%%%%%%%%%%%%%%%%%%%%%%%%%%%
\subsection{Numerical tests of the universal square-root singularity\label{singularity}}
The calculation in Sec.~\ref{sec:scaling} shows that the approach to the instability has critical features.
Here we examine the numerical ramifications of this criticality. We note that when the strain is increased two real eigenvalues can collide to produce a complex conjugate pair, but the opposite can also happen as long as we increase the strain subject to overdamping; a pair of complex conjugate pair can give rise to two real eigenvalues.
The square root singularity applies to both transitions.

\begin{figure}
%\vskip 0.4 cm
  \includegraphics[width=0.35\textwidth]{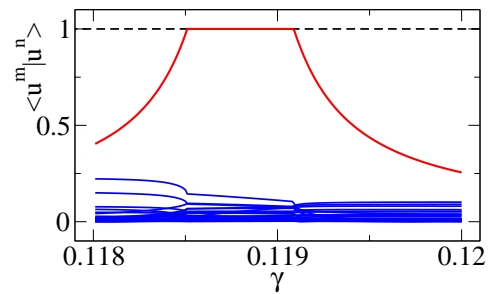}\hspace{1mm}
  \caption{
    Scalar products between the eigenvector $|{u}^{m}\rangle$ of a fixed mode $m$ (here $m=25$) and all the other eigenvectors (total 29) over shear strain $\gamma$ for a system of 10 disks. The dashed line is added to guide the eye.
  }
  \label{Fig:scalar}
\end{figure}

\begin{figure}
  \includegraphics[width=0.35\textwidth]{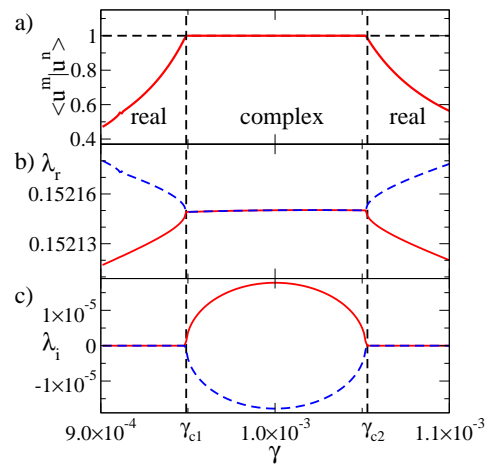}
  \caption{ (a) the scalar product of two eigenvectors for a system of $N=500$ and $\mu=10$.
  The two modes are complex conjugate in the strain interval varying between $\gamma_{c1}$ and $\gamma_{c2}$.
  (b) The real and (c) the imaginary term of the two corresponding eigenvalues consistently show that these are complex conjugates for strain values in between $\gamma_{c1}$ and $\gamma_{c2}$.
  \label{Fig:complexpair}}
\end{figure}

A good indicator for the birth and presence of two complex conjugate eigenfunctions over some range of strain values is provided by the scalar product of the two colliding modes.
The scalar product of two eigenvectors is defined as
$\langle{u}^{m}| {u}^{n}\rangle = |\sum_{j=1}^{3N} (u_{r;j}^{m}+iu_{i;j}^{m})\cdot(u_{r;j}^{n}+iu_{i;j}^{n})|$. $\langle{u}^{m}| {u}^{n}\rangle$ varies in the interval $[0,1]$. At the critical value $\gamma_c$ the scalar product reaches unity. This is quite obvious since at the critical point also the eigenfunctions become complex conjugates. We show this first in Fig.~\ref{Fig:scalar} for a system of 10 disks and in Fig.~\ref{Fig:complexpair} for 500 disks. Note that  the eigenvectors of a real non-symmetric matrix are linearly independent, but in general not orthogonal.

\begin{figure}
  \includegraphics[width=0.42\textwidth]{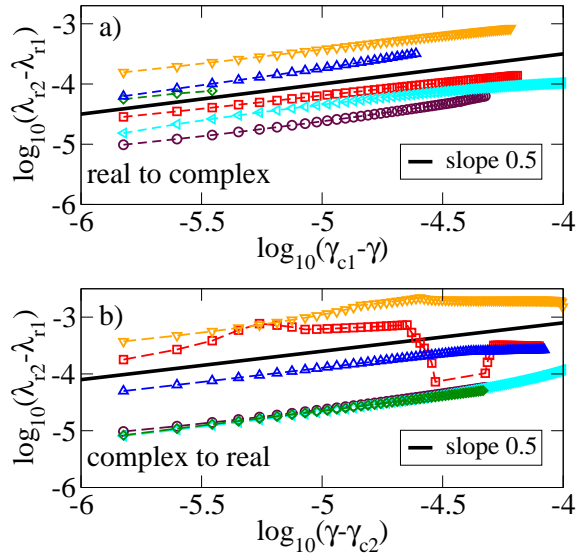}
  \caption{
  Critical behaviour of the difference between two real eigenvalues as they collide as $\gamma$ approaches $\gamma_{c1}$ (a),
  or depart as $\gamma$ overcomes $\gamma_{c2}$. The different curves refer to events occurring in different samples with $N$ (500,1000) and $\mu$ (0.5,10). The solid lines are straight lines with slope 0.5.
  }
  \label{fig:collision}
\end{figure}

A direct test of the universality predicted in the last subsection is provided by
the difference between the two real eigenvalues. The data are consistent with
$|\lambda_r^{1}-\lambda_r^{2}| \propto (\gamma_c-\gamma)^{x}$, and $x \simeq 0.5$.
This is shown in Fig.~\ref{fig:collision}.
%%%%%%%%%%%%%%%%%%%%%%%%%%%%%%%%%%%%%%%%%%%%%%%%%%%%%%%%%%%%%%%%%%%%%%5
%%%%%%%%%%%%%%%%%%%%%%%%%%%%%%%%%%%%%%%%%%%%

%%%%%%%%%%%%%%%%%%%%%%%%%%%%%%%%%%%%%%%%
\section{The development of the instability~\label{develop}}
\subsection{Linear response regime\label{secspiral}}
To investigate the development of an instability, we consider the Newtonian (no damping) evolution
of a configuration having a pair of complex conjugate eigenvalues.
As long as the deviations from mechanical equilibrium are sufficiently small, the eigenvalues and the eigenfunctions of $\B J$ furnish an accurate prediction for the developing instability.
To see this denote the complex mode $m$ of ${\bf J}$ whose complex eigenvalue $\lambda^m = \lambda_r + i \lambda_i$ as
\begin{equation}
  |u^m\rangle = |u^m_r\rangle + i |u^m_i\rangle.
\end{equation}
Since the dimension of ${\bf J}$ is $3N\times 3N$ then $|u^m_r\rangle$ and similarly $|u^m_i\rangle$ have $3N$ components.
As
\begin{equation}
  {\bf J} |u^m\rangle = \lambda^m | u^m\rangle
\end{equation}
we can rewrite the above equation in the form:
\begin{eqnarray}
  {\bf J}|u^m_r\rangle &=& \lambda_r |u^m_r\rangle - \lambda_i |u^m_i\rangle \\
  {\bf J}|u^m_i\rangle &=& \lambda_i |u^m_r\rangle + \lambda_r |u^m_r\rangle.
\end{eqnarray}
In matrix notation this can be expressed as follows:
\begin{equation}
  {\bf J} \begin{bmatrix}{\bf u}^m_r &{\bf u}^m_i\end{bmatrix}_{3N\times 2} = \begin{bmatrix}{\bf u}^m_r & {\bf u}^m_i\end{bmatrix}_{3N\times 2} \times  \begin{bmatrix}\lambda_r & \lambda_i \\ -\lambda_i & \lambda_r \end{bmatrix}_{2\times 2}.
\end{equation}

\begin{figure}
    \includegraphics[width=0.4\textwidth]{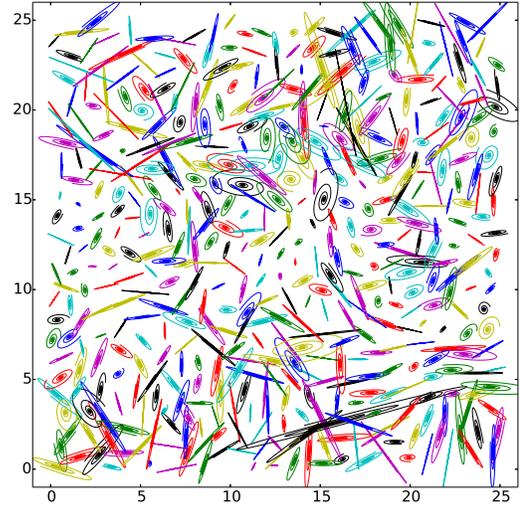}
  \caption{A typical image of spiral trajectories of 500 disks in linear response regime. Here actual particle displacements are amplified by a factor $10^6$.}
    \label{spiral}
\end{figure}

\begin{figure}
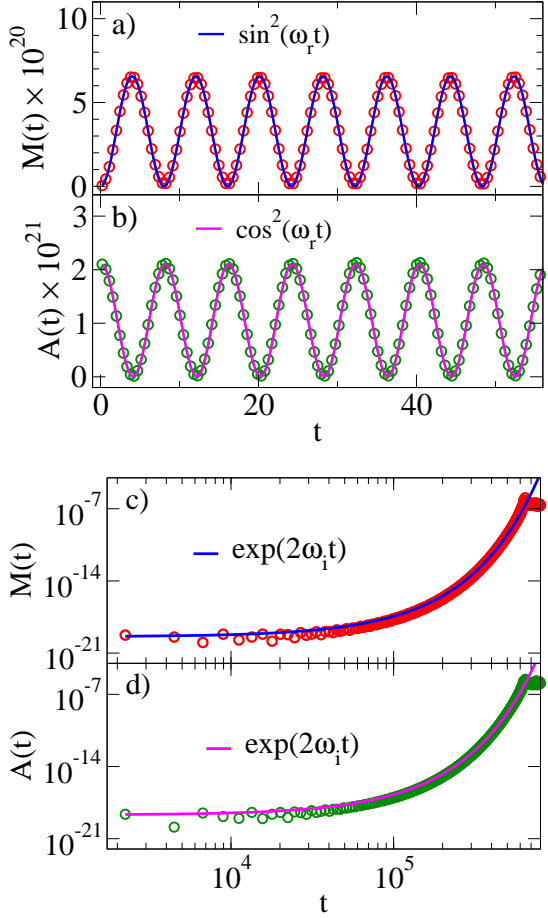

  \includegraphics[width=0.4\textwidth]{{Follow.Osc.Fig6a}.eps}
  \vskip 0.4 cm
  \includegraphics[width=0.4\textwidth]{{Follow.Osc.Fig6b}.eps}
  \caption{(a) $M(t)$ and (b) $A(t)$  during the early stage of Newtonian dynamics, after setting the velocities (both translation and rotation) of 500 disks along the resultant vector $\ddot{\bf u}^m$, Eq.~(\ref{equm}). Both quantities display sinusoidal motion with the expected real eigenfrequency $\omega_r$. (c) $M(t)$ and (d) $A(t)$ grow exponentially following the form $\sim \sin^2(\omega_rt+\phi)\exp(2\omega_it)$, where $\phi$ is the initial phase, zero for $M(t)$ and $\pi/2$ for $A(t)$. The growth stops typically at $M(t) \sim 10^{-7}$ when either a few contacts break or a few new contacts get created. }
    \label{perturbed}
\end{figure}

The right hand side of the above equation produces a dynamic matrix of dimension $3N\times 2$:
%\begin{equation}
  %\ddot{\bf u} \equiv  \begin{bmatrix}{\bf u}^m_r & {\bf u}^m_i\end{bmatrix}_{3N\times 2} \times  %\begin{bmatrix}\lambda_r & \lambda_i \\ -\lambda_i & \lambda_r \end{bmatrix}_{2\times 2} \\
    %= \begin{bmatrix}\ddot{{\bf u}}^1 & {\ddot{\bf u}}^2\end{bmatrix}_{3N\times 2} \ .
%\end{equation}
\begin{equation}
  \begin{bmatrix}\ddot{{\bf u}}^1 & {\ddot{\bf u}}^2\end{bmatrix}_{3N\times 2} =  \begin{bmatrix}{\bf u}^m_r & {\bf u}^m_i\end{bmatrix}_{3N\times 2} \times  \begin{bmatrix}\lambda_r & \lambda_i \\ -\lambda_i & \lambda_r \end{bmatrix}_{2\times 2} \\
     \ .
\end{equation}
In particular, under the operation of this matrix, the resultant of
$\ddot{{\bf u}}^1$ and $\ddot{{\bf u}}^2$, i.e.
\begin{equation}\label{equm}
  \ddot{\bf u}^m = \ddot{{\bf u}}^1 + {\ddot{\bf u}}^2 \
\end{equation}
is subject to a rotation-scaling operation, and is therefore expected to predict a
spiral trajectory with exponentially increasing speed.
To demonstrate this, we consider a system in equilibrium and set the velocities
of the particles (both translational and rotational) along the direction fixed by
$\ddot{\bf u}^m$, and then follow the evolution of the system.
Fig.~\ref{spiral} clearly shows that particles display the expected spiral motion.
The corresponding evolution of the mean square displacement of the center of mass
$M(t)$ and the mean square change in the angular coordinate $A(t)$ can also be determined.
Denoting
\begin{eqnarray}
&&\Delta r^x_i(t) \equiv r^x_i(t) -r^x_i(t=0)\ , \nonumber\\
&&\Delta r^y_i(t) \equiv r^y_i(t) -r^y_i(t=0)\ , \nonumber\\
&&\Delta \theta_i(t) \equiv \theta_i(t)- \theta_i(t=0)\ ,
\end{eqnarray}
we define
\begin{eqnarray}
M(t) &\equiv &\frac{1}{N} \sum_i^N [(\Delta r^x_i(t))^2 + (\Delta r^y_i(t))^2]\ , \nonumber\\
A(t) & \equiv & \frac{1}{N} \sum_i^N (\Delta \theta_i(t))^2 \ .
\end{eqnarray}
According to Eq.~(\ref{growth}) the system should behave as $M(t)\propto A(t) \propto  e^{2 \omega_i t} \sin^2(\omega_r t)$.
As shown in Fig.~\ref{perturbed} both quantities display sinusoidal motion since the beginning of the Newtonian dynamics with the expected frequency $\omega_r$ and the expected exponential growth $e^{2\omega_i t}$.
%%%%%%%%%%%%%%%%%%%%%%%%%%%%%%%%%%%%%
%%%%%%%%%%%%%%%%%%%%%%%%%%%%%%%%%%%%%%%%%%%%

\subsection{The Nonlinear regime\label{nonlinear}}
%%%%%%%%%%%%%%%%%%%%%%%%%%%%%%%%%%%%%%%%%
\begin{figure}
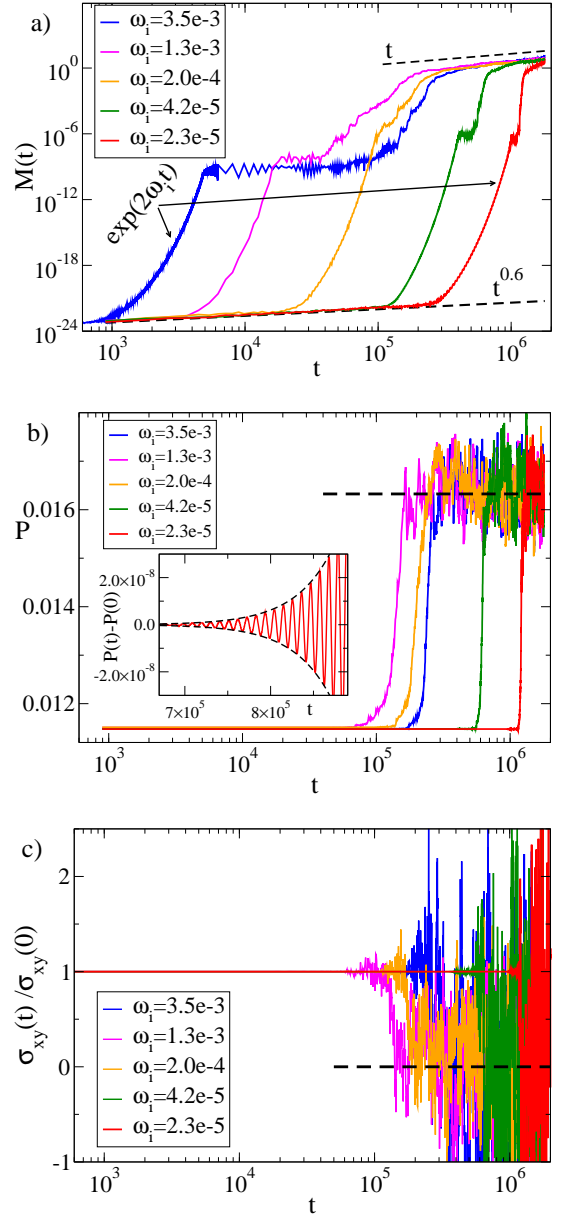

  \includegraphics[width=0.4\textwidth]{{Follow.Osc.Fig7a}.eps}
  \vskip 0.4 cm
  \includegraphics[width=0.4\textwidth]{{Follow.Osc.Fig7b}.eps}
  \vskip 0.4 cm
  \includegraphics[width=0.4\textwidth]{{Follow.Osc.Fig7c}.eps}
  \caption{Long-time fate of the oscillatory instability evolving with a newtonian dynamics for $N=500$ and $\mu=10$.
  Different lines correspond to different initial configurations, identified by their complex eigenvalue with the largest imaginary component (growth rate).
   (a) Mean square displacement $M(t)$, (b) pressure $P$ and (c) shear stress $\sigma_{xy}$.
   Inset of (b) $P$ (solid line) in the linear response regime for $\omega_i\sim 2.3\times 10^{-5}$.
   Two dashed lines are $\pm \exp(\omega_it)$ functions.
  }
    \label{long1}
\end{figure}

%%%%%%%%%%%%%%%%%%%%%%%%%%%%%%%%%%%%%%%%%%%%
When the instability develops sufficiently, the system exits from the linear response regime.
It is of interest to consider how this happens, and what is the long time fate of the instability.
We address this point following the evolution of a number of unstable configurations
in which the linear instability is associated with complex eigenvalues with very different frequencies.
It is noteworthy that while these initial configurations usually have more than a single pair of complex eigenfunctions
the dynamics in the linear regime is dominated by the complex mode with the largest $\omega_i$,
that defines the growth rate of the instability.
%The linear regime of the purely exponential growth ends when the neighbor list begins to change. Typically a few contacts break or a few new contacts get created. In the present simulation this happens typically at $M(t)$ value $\sim 10^{-7}$.
%It is interesting to follow the development of the instability.
%After the exponential growth the dynamics slows down, and one observes a plateau in $M(t)$. Without damping, the plateau is followed by a new instability which is accompanied by a marked increase in the pressure, and drop of the shear stress. This is a system spanning mechanical failure which is associated with a major change in the configuration.
We illustrate the evolution of these different systems in Fig.~\ref{long1}, which reports the time evolution
of the mean square displacement (a), of the pressure (b), and of the shear stress (c).
From these figure, we identify in the dynamics the following regimes:
\begin{enumerate}
 \item at short times, the dominant effect of the oscillatory instabilities results in an exponential growth in $M(t)\sim \sin^2(\omega_rt)\exp(2\omega_it)$. Similarly, as illustrated in the inset of Fig.~\ref{long1}b, the pressure $P$ oscillates around its initial value with the oscillation amplitude increasing exponentially as $P(t)-P(0) \sim \sin(\omega_rt)\exp(\omega_it)$. The dashed lines in the inset illustrate the envelope $\pm \exp(\omega_it)$, where $\omega_i\sim2.3\times10^{-5}$.
   The shear stress is also oscillating as the pressure around a constant value in the linear response regime, already shown explicitly in~\cite{19CGPP}.

 \item the exponential growth is interrupted when $M(t) \sim 10^{-7}$, which occurs at a characteristic time proportional to $\omega_i^{-1}$.
 At this time the two unstable complex conjugate modes annihilate, as a consequence of the breaking of few contacts (see below) and  $M(t)$ enters a plateau regime. In this plateau regime, $P$ and $\sigma$ exhibit enhanced fluctuations, but no noticeable change in their mean values.
 \item the plateau regime is interrupted by a fast growth of $M(t)$ driven by a second instability, which is triggered by the emergence of a number of complex eigenvalues. Afterwards, the system enters a diffusive regime.
This second instability leads to a sharp increase in the pressure, and to a dramatic drop in shear stress $\sigma_{xy}$.
 \end{enumerate}

\begin{figure}
  \includegraphics[width=0.4\textwidth]{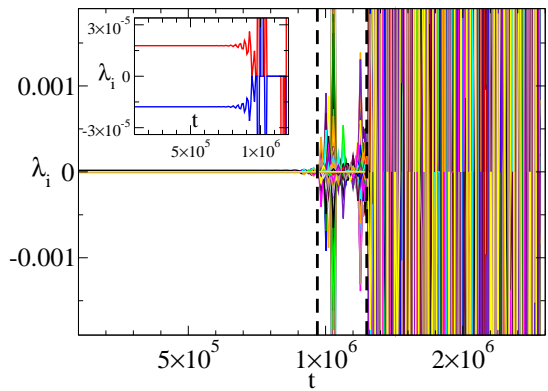}
  \caption{Evolution of the $3N = 1500$ imaginary component of the eigenvalues $\lambda_i$ during a Newtonian dynamics. The initial configuration contains one single pair of complex eigenpair with $\omega_i \sim 2.3 \times 10^{-5}$, whose evolution is illustrated in the inset. The left vertical dashed line signifies the end of existing complex eigenpair that further coincides with the end of exponential growth in $M(t)$. The right vertical dashed line signifies the birth of many complex eigenpairs that initiates the second instability in $M(t)$, see Fig.~\ref{long1}.
  }
    \label{long2}
\end{figure}

Fig.~\ref{long2} illustrates how the development of the instability is related to the change in the spectrum.
The inset follows the evolution of the imaginary part of the complex eigenvalues dominating at short time, and show that this vanishes when the exponential grow phase ends, for the considered $\omega_i\sim2.3\times10^{-5}$.
The main panel follows the evolution of the imaginary part of all eigenvalues, and demonstrates that the secondary instability is triggered by the emergence of several unstable modes. The two consecutive vertical black dashed lines in Fig.~\ref{long2} represent the end of the exponential growth and the beginning of the second instability.

%%%%%%%%%%%%%%%%%%%%%%%%%%%%%%%%%%%%%%%%%%%%

%%%%%%%%%%%%%%%%%%%%%%%%%%%%%%%%%%%%%%%%%%%%
\begin{figure}
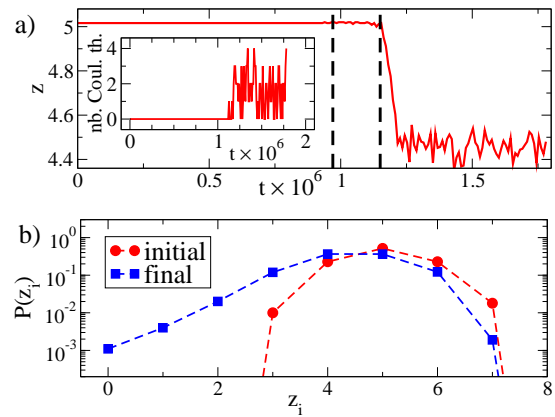

  \includegraphics[width=0.4\textwidth]{{Follow.Osc.Fig9a}.eps}\\
  \vskip 0.25 cm
  \includegraphics[width=0.4\textwidth]{{Follow.Osc.Fig9b}.eps}
    \caption{Newtonian dynamics. (a) Number of contacts $z$ on average a disk has exhibiting a sharp drop at second instability shown in Fig.~\ref{long1}. The dashed lines correspond to (i) the collapse of the initial complex pair that coincides with a few make or break of contacts and (ii) birth of many complex modes coincides with the verge of the drop in contacts, see Fig.~\ref{long2}. (inset) Number of contacts are at Coulomb threshold. (b) The probability distribution function for a disc to have $z$ contacts before and after the drop.
  }
    \label{long3}
\end{figure}
%%%%%%%%%%%%%%%%%%%%%%%%%%%%%%%%%%%%%%%%%%%%

The comparison of Fig.~\ref{long2} and Figs.~\ref{long1}b,c clarifies that the effect of a dying complex mode might be eventful from a mechanical viewpoint. On the one hand it activates many new complex modes. On the other hand the shear stress $\sigma_{xy}$ drops almost to zero.
Regarding the structure of the system, we observe in Fig.~\ref{long3}(a) that the end of the exponential growth is associated to the destruction of few contacts. In addition a few contacts achieve the Coulomb threshold (inset).
A significant restructuring of the contact network only occurs during the second instability, which is associated with a drop in the number of contacts between grains. This drop leads to a significant change in the probability distribution of the number of contacts per particle, $P(z_i)$ due to the emergence of particles with few contacts, as in Fig.~\ref{long3}(b).

\begin{figure}
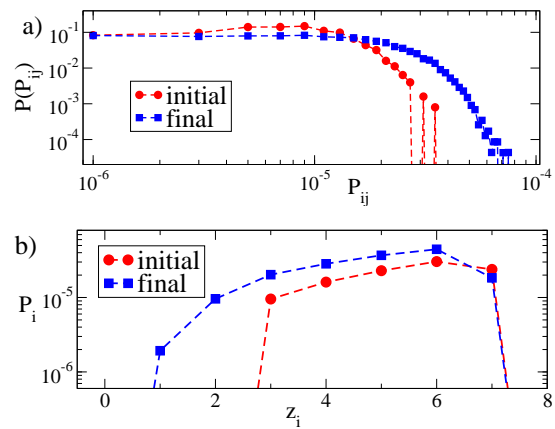

  \includegraphics[width=0.4\textwidth]{{Follow.Osc.Fig10a}.eps}
  \vskip 0.25 cm
  \includegraphics[width=0.4\textwidth]{{Follow.Osc.Fig10b}.eps}
  \caption{Newtonian dynamics. (a) the distribution of binary pressure $P_{ij}$ and (b) the pressure $P_i$ exerted on each particle having contacts $z$ before and after the nonlinear dynamics, see text for details.
}
    \label{pressure}
\end{figure}

We finally clarify that the sharp increase in the pressure observed during the second instability
is associated to the fluidization of the system. To this end, we introduce a per-particle pressure defined as $P_i=\frac{1}{4L^2}\sum_{j=1}^{N}{\bf r}_{ij}\cdot{\bf F}_{ij}$ such that $P=\sum_{i=1}^NP_i$. Fig.~\ref{pressure}a illustrates the distribution of $P_{ij}={\bf r}_{ij}\cdot{\bf F}_{ij}$ at time $t=0$, and at a later time after the secondary instability.
Clearly, in the final state the pressure distribution has a longer tail, consistent with the observation of a larger average pressure value.
Fig.~\ref{pressure}b illustrates the average value of $P_i$ of the particles having $z$ contacts.
In the final configuration, the pressure is higher for all values of $z$, as a consequence of the existence of large particle deformation associated to the flow of the system.

%%%%%%%%%%%%%%%%%%%%%%%%%%%%%%%%%%%%%%%%%%%%%%%%%%%%%%%%%%%%%%%%%%%%%%%%%%%%%%%%%%%%%%%%%%%%%%%%%
\section{The effects of damping\label{damp}}
While we have investigated the development of the oscillatory instability solving Newton's equations with no dissipation, in most real system applications there is some form of damping.
It can be due to the water and grit in a fault, to the interparticle interaction,
or to additional friction with a confining substrate.
It is therefore important to ask if and how damping affects the
observed instabilities. In this section we show that damping tames but does not kill the instabilities.
In particular, the instabilities associated with a small exponential growth rate (and hence emerging from the collision of low frequency modes) are more affected by the damping.

For our investigation we consider a starting configuration obtained from our athermal quasistatic simulations,
which has a complex eigenpair with eigenfrequency $\omega_r \pm i\omega_i= 0.3900643\pm i 0.0000228$.
We then evolve the system according to a Newtonian dynamics, where in Eq.~(\ref{Newton2a}) we add a damping force $-\tilde m_i\eta_0 \dot {\B r}_i$ on the right hand side.
%%%%%%%%%%%%%%%%%%%%%%%%%%%%%%%%%%%%%%%%%%%%
\begin{figure}
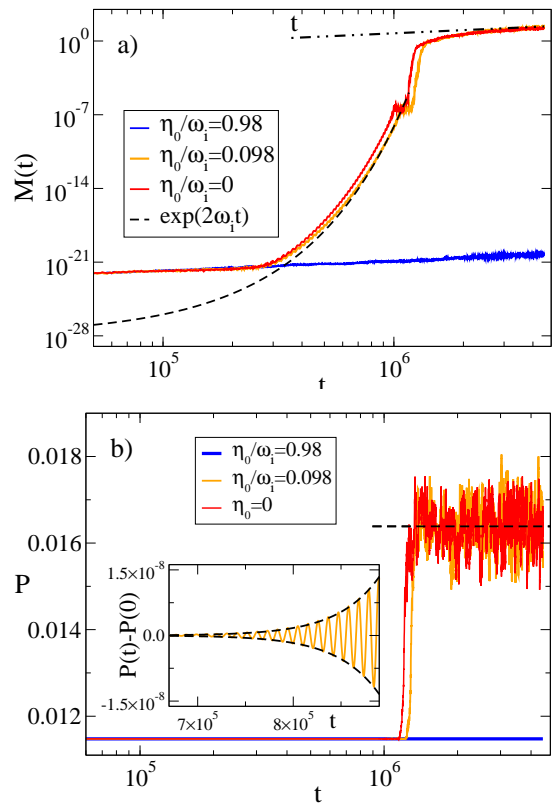

  \includegraphics[width=0.4\textwidth]{{Follow.Osc.Fig11a}.eps}
  \vskip 0.2 cm
  \includegraphics[width=0.4\textwidth]{{Follow.Osc.Fig11b}.eps}
  \caption{The effect of damping on complex eigenfrequency. (a) Mean square displacement $M(t)$ and (b) pressure $P$ display the same characteristic behaviors as for zero damping force when the damping frequency $\eta_0$ is smaller than the complex eigenfrequency $\omega_i$.
  }
    \label{damping}
\end{figure}
%%%%%%%%%%%%%%%%%%%%%%%%%%%%%%%%%%%%%%%%%%%%
Fig.~\ref{damping} shows that the evolution of this damped dynamics depends on how
the damping rate $\eta_0$ compares with the instability growth rate $\omega_i$,
focussing on the time dependence of the mean square displacement (a) and of the pressure (P).
For  $\eta_0 \gtrsim \omega_i$, the exponential growth is suppressed.
For $\eta_0 < \omega_i$, the dynamics has all the same features we have found in Fig.~\ref{long1}.
As an example, in the inset of the lower panel we show that the pressure $P$ oscillates around
its initial value while the oscillation amplitude increases exponentially as $P(t)-P(0) \sim \sin(\omega_rt)\exp(\omega_it)$,
as observed in the absence of damping in Fig.~\ref{long1}b.
Overall, this analysis clarifies that damping suppresses the instabilities with a small growth rate $\omega_i$,
not those corresponding to a large growth rate $\omega_i$.

\section{How generic is the oscillatory instability?}
 \label{howgen}
 Having presented the oscillatory instability and its ability to self-amplify small perturbations, it is important to ask how generic is this instability. In particular, is the oscillatory instability a consequence of the structure of the Hertz-Mindlin model as interpreted for example by Cundall and Strack \cite{79CS}, or is it expected to appear in any dynamical model of frictional disks. The answer to this question depends on what do we mean by ``a model" of frictional disks. During the last few decades the granular community has accepted that assemblies of disks can be described by coarse-grained coordinates, for example the coordinates $\B r_i$ of the centers of mass and $\theta_i$ for the angle of the disk (with reference to an initial angle $\theta_i(t=0)$). From the knowledge of these coordinates one writes down a physical model of the normal and tangential forces at each contact between neighboring disks. The other possibility is to track the exact micro-dynamics at each contact, paying attention to its leading and trailing edges, and taking into account the roughness of the contact and the plastic and elastic events that take place. This approach is not realistic for large assemblies of frictional disks, and even less so for arbitrarily shaped granules. Thus, if one follows the approach of coarse-grained coordinates, the options are somewhat limited.
 Imagine for example that we decide to
 use the harmonic-type normal interaction and delete $\sqrt{\delta_{ij}}$ term from the tangential force in Eq.~(\ref{Min}).
 Then on the face of it, the forces could be derived from a Hamiltonian of the form
 \begin{equation}
 U(\delta_{ij},t_{ij}) = \frac{1}{2}k_n(\delta_{ij})^{2} + \frac{1}{2}k_t(t_{ij})^2 \ .
 \label{hamil}
 \end{equation}
 Will this suffice to eliminate the oscillatory instability? The answer is no, as long as we do not
 remove the Coulomb law and the requirement of smoothness of forces to allow derivatives. Any smoothing
 of the type done in Eq.~(\ref{Ft}) brings back the dependence of the tangential force on the normal
 force and makes the system non-Hamiltonian.
 %%%%%%%%%%%%%%%%%%%%%%%%%%%%%%%%%%%%%%%%%%%%
\begin{figure}
  \includegraphics[width=0.4\textwidth]{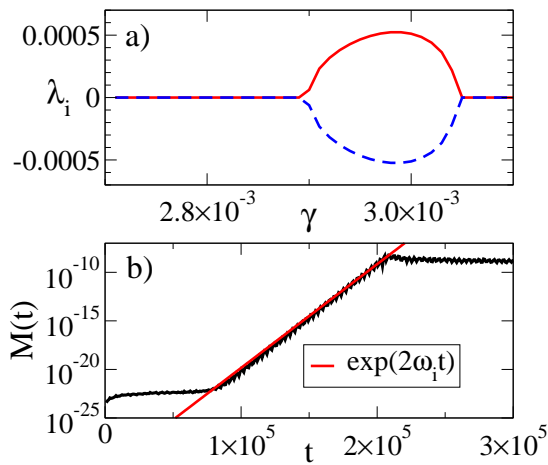}
  \caption{(a) The typical characteristic of a complex eigenpair during an AQS simulation is recovered for a granular assembly of 500 particles where the normal interaction force is a harmonic function $k_n\delta$, and the tangential force is a smooth and continuous function $k_t[1+(t/t^*) - (t/t^*)^2]t$, and at Coulomb threshold $t=t^*$, it is $\mu k_n\delta$ with $\mu=10$.
    (b) Newtonian dynamics displays an exponential growth in mean square displacement as expected for a system initially having a complex eigenpair.
  }
    \label{harmonic}
\end{figure}
%%%%%%%%%%%%%%%%%%%%%%%%%%%%%%%%%%%%%%%%%%%%

 To demonstrate this we have rerun our dynamics after removing the term $\sqrt{\delta_{ij}}$ from the tangential force in Eq.~(\ref{Min}), but keeping the smoothing of this force at the Coulomb threshold. The  resulting oscillatory instability is displayed in Fig.~\ref{harmonic}. As explained, the system still
 does not have a Hamiltonian structure, and the oscillatory instability remains generic. To drive home
 the message even further, we now remove the requirement of a smooth tangential force at the Coulomb
 threshold, allowing the tangential force to reach the limit in an abrupt way. Now the oscillatory
 instability disappears, since the Hamiltonian (\ref{hamil}) remains now valid all the way
 to the Coulomb threshold. The price is of course that the Jacobian matrix $\B J$ does not exist
 since the tangential force is not derivable at the threshold.

 The above findings lead us to an interesting juncture for studying frictional properties in granular materials.
 The Hamiltonian models which oversimplify the role of deformation between two particles on static friction do not possess the oscillatory instabilities.
 In reality, the deformation between contacts is complex as found by various experiments and atomistic simulations, and a direct intrusion of overlap distance in the tangential force is the culmination of this fact. The overall dynamics becomes non-Hamiltonian that further displays oscillatory instabilities with a cost of no guarantee in energy conservation.
At this point one depends on experiments to verify or refute the existence of this micromechanics in granular materials.

\section{Summary and the Road Ahead}
\label{summary}
It appears from our discussion that oscillatory instabilities should be expected as a generic feature present in physical system whose dynamics are not derivable from a Hamiltonian.
This novel instability was recently reported for frictional assemblies of disks \cite{19CGPP}.
In this article, we have explored further the characteristics and the consequences of such instabilities on similar frictional granular systems.

We have shown that the birth and death of an oscillatory instability upon athermal quasi-static shear are a critical phenomenon. A pair of complex eigenpair gets born when two real eigenmodes align to one another, the relative difference between the two real eigenvalues vanishes with shear strain following a power law function with exponent 1/2. Similarly, the death of the complex eigenpair takes place when the alignment between the two eigenvectors is lost. The relative difference between the two real eigenvalues gradually increases following the same power law as a function of the strain.

With Newtonian dynamics the presence of a complex eigenpair generates instabilities where the oscillatory amplitudes increase exponentially. During this linear response regime particles move in spiral trajectories whose directions are determined by the two complex eigenvectors of the corresponding eigenpair. We have presented a formalism describing the direction of the spiral trajectories in the real plane; it is determined by the two complex eigenvectors. The linear response regime comes to an end when the oscillatory amplitude becomes so large that it breaks or makes a few pair contacts.
This has a dramatic consequences;  this triggers a second instability due to which the system goes through a structural change. The number of contacts of a particle becomes broadly distributed resulting in an increase of pressure. The dynamics becomes diffusive, helping the system to dissipate the shear stress.

The above phenomena are generic for a whole range of complex eigenvalues and even in the presence of damping force as long as the damping frequency is smaller than the imaginary eigenfrequency.

Probably the most important conclusion is that we have uncovered here a mechanism for a giant amplification of small perturbations.
This finding may have further consequences in the context of a number of physics problems that involve frictional interactions.
One fundamental question is how static friction turns into dynamical friction~\cite{Urbakh2004,Braun2009,Tromborg2011}.
So far it is not quite clear what is the precise instability that allows two compressed interfaces
to start moving with respect to one another.
It is worthwhile in the coming future to examine whether the type of instability discussed above may be responsible for setting in this interesting transition.
Another context of interest is that of remote triggering.
It is known that one earthquake can induce another earthquake far away~\cite{06FB,00BKK,14BE}.
Since strong seismic waves are strongly damped, only weak perturbation can
travel a long distance.
So the phenomenon of remote triggering requires a mechanism for self amplification.
While we are fully aware that geological faults are very much more complex than assemblies of frictional disks, it appears highly worthwhile to explore the relevance of the kind of self amplification of small perturbation discussed above also to the problem of remote triggering.
It is certainly on our agenda for the near future.

\acknowledgments
This work had been supported in part by the ISF-Singapore exchange program and by
the US-Israel Binational Science Foundation. We thank Jacques Zylberg and Yoav Pollack
for useful discussions and exchanges at the early stages of this project.
JC and MPC acknowledge NSCC Singapore for granting the computational facility under project 12000621.

\appendix
\section{Calculation of the operator $\B J$~\label{Jacobian}}
The Jacobian operator ${\B J}$, which is the dynamical response of the system, represents the derivative of the forces and of the torques acting on the particles with respect
to all the degrees of freedom. The interaction forces used in this study is recalled in Sec.~\ref{sec:forces}, then the expression of the tangential displacement and its derivative (Sec.~\ref{sec:tangential}) and the derivatives of forces and torques with respect to generalized coordinates (Sec.~\ref{sec:J}) are stated. Finally, we show the expressions for all the components of ${\B J}$ (Sec.~\ref{sec:Jcomponents}) and how these components are arranged as a matrix (Sec.~\ref{sec:Jarrangement}).

%%%%%%%%%%%%%%%%%%%%%%%%%%%%%%%%%%%%%%%%%%%%%%%%%%%%%%%%%%%%%%%%%%%%%%%%%%%%%%%%%%%
\subsection{Interaction force~\label{sec:forces}}
In our simulation, a pair of granular particles interacts when they overlap. The overlap distance $\delta_{ij}$ is measured as
\begin{equation}
  \delta_{ij}=\sigma_i+\sigma_j-r_{ij},
  \label{Eq:delta}
\end{equation}
where $r_{ij}$ is the center-to-center distance of a pair-$i$ and $j$, and $\sigma_i$ is the radius of particle-$i$. The pair vector ${\B r}_{ij}$ is defined as
\begin{equation}
  {\B r}_{ij}=\B r_i - \B r_j.
  \label{Eq:rij}
\end{equation}
The pair-interaction force ${\B F}_{ij}$ has two contributions. ${\B F}^{(n)}_{ij}$ is the force acting along the normal direction of the pair $\hat{r}_{ij}$, and ${\B F}^{(t)}_{ij}$ is the force acting along the tangential direction of the pair ${\hat t}_{ij}$.
The normal force is Hertzian:
\begin{equation}
  {\B F}^{(n)}_{ij}= k_n\delta^{3/2}_{ij} {\hat r}_{ij},
  \label{Eq:Fn}
\end{equation}
where $k_n$ is the force constant with dimension: Force per length${}^{3/2}$.
The tangential force ${\B F}^{(t)}_{ij}$ is a function of both the overlap distance $\delta_{ij}$ and the tangential displacement ${\B t}_{ij}$.
We have modified the standard expression for ${\B F}^{(t)}_{ij}$ and included a few higher order terms of $t_{ij}$ (i.e., $|{\B t}_{ij}|$) such that the derivative of the force function $F^{(t)}_{ij}$ with respect to tangential distance $t_{ij}$ becomes continuous and it goes to zero smoothly.
We use the following form:
\begin{equation}
  \label{Eq:Fs}
  \begin{split}
    {\B F}^{(t)}_{ij} & = -k_t\delta^{1/2}_{ij}\left[1 + \frac{t_{ij}}{t^*_{ij}} - \left(\frac{t_{ij}}{t^*_{ij}}\right)^2 \right] t_{ij}{\hat t}_{ij}\\
    & = -k_t \delta^{1/2}_{ij} t^*_{ij}{\hat t}_{ij}, \ \ \ \textrm{if}\ \ k_t\delta^{1/2}_{ij}t_{ij} > \mu |{\B F}^{(n)}_{ij}|,
  \end{split}
\end{equation}
where $k_t$ is the tangential force constant. Its dimension is force per length${}^{3/2}$. $t^*_{ij}$ is the threshold tangential distance:
\begin{equation}
t^*_{ij} = \mu \frac{k_n}{k_t} \delta_{ij},
\label{Eq:s*}
\end{equation}
where $\mu$ is the friction coefficient, a scalar quantity, which essentially determines the maximum strength of the tangential force with respect to the normal force at a fixed overlap $\delta_{ij}$. The derivative of $F^{(t)}_{ij}$ with respect to $t_{ij}$ vanishes at $t^*_{ij}$, as it turns out
\begin{equation}
 \label{Eq:dFsds}
  \begin{split}
    \frac{\partial F^{(t)}_{ij}}{\partial t_{ij}} & = k_t\delta^{1/2}_{ij}\left[1 + 2\frac{t_{ij}}{t^*_{ij}} - 3\left(\frac{t_{ij}}{t^*_{ij}}\right)^2 \right] \\
    & = 0, \ \ \ \textrm{if}\ \ k_t \delta^{1/2}_{ij}t_{ij} > \mu |{\B F}^{(n)}_{ij}|.
  \end{split}
\end{equation}

{\bf We stress here that the above forces imply a non Hamiltonian dynamics.} That is, there is not
a function $U(\delta,t)$ such that $F^{(n)} = -\frac{\partial U}{\partial \delta}$ and
$F^{(t)} = -\frac{\partial U}{\partial t}$.

%%%%%%%%%%%%%%%%%%%%%%%%%%%%%%%%%%%%%%%%%%%%%%%%%%%%%%%%%%%%%%%%%%%%%%%%%%%%%%%%%%%

%%%%%%%%%%%%%%%%%%%%%%%%%%%%%%%%%%%%%%%%%%%%%%%%%%%%%%%%%%%%%%%%%%%%%%%%%%%%%%%%%%%
\subsection{Tangential displacement\label{sec:tangential}}
The tangential force is a function of both $\B t_{ij}$ and ${\B r_{ij}}$. The derivative of this force thus includes the derivative of the two latter quantities. Here we evaluate these derivatives
using the chain rule.

The derivative of tangential displacement ${\B t}_{ij}$ with respect to time $t$ is
\begin{equation}
  \frac{\mathrm d {\B t}_{ij}}{\mathrm d t} = {\B v}_{ij} - {\B v}^n_{ij} + \hat{r}_{ij}\times(\sigma_i{\B \omega}_i + \sigma_j{\B \omega}_j),
  \label{Eq:dsdt}
\end{equation}
where ${\B v}_{ij}={\B v}_i-{\B v}_j$ is the relative velocity of pair-$i$ and $j$. ${\B v}^n_{ij}$ is the projection of ${\B v}_{ij}$ along the normal direction $\hat{r}_{ij}$. ${\B v}_{ij} - {\B v}^n_{ij}$ is the tangential component of the relative velocity.
${\B \omega}_i$ and ${\B \omega}_j$ are the angular velocity of $i$ and $j$, respectively.
In differential form, the above equation reads:
\begin{equation}
  \mathrm d {\B t}_{ij} =  \mathrm d{\B r}_{ij} -  (\mathrm d{\B r}_{ij}\cdot\hat{r}_{ij})\hat{r}_{ij} + \hat{r}_{ij}\times(\sigma_i\mathrm d{\B \theta}_i + \sigma_j\mathrm d{\B \theta}_j),
  \label{Eq:ds}
\end{equation}
where $\mathrm d{\B \theta}_i$ is the angular displacement of $i$ which follows the relation: $\mathrm d{\B \omega}_i = \frac{\mathrm d {\B \theta}_i}{\mathrm d t}$.

Here on, we assume the two-dimensional ({$\bf 2D$}) system.
Therefore, $\omega_i$, and so $\theta_i$, only have one component along $\hat{z}$, perpendicular to the xy plane, and $\hat{r}_{ij}\times\mathrm d{\B \theta}_i = \mathrm d \theta_i (y_{ij}\hat{x}-x_{ij}\hat{y})/r_{ij}$. This allows to write Eq.~(\ref{Eq:ds}) as
\begin{equation}
  \mathrm d t^\alpha_{ij} =  \mathrm d r^\alpha_{ij} -  (\mathrm d{\B r}_{ij}\cdot\hat{r}_{ij})\frac{r^\alpha_{ij}}{{r}_{ij}} + (-1)^\alpha(\sigma_i\mathrm d \theta_i + \sigma_j\mathrm d\theta_j)\frac{r^\beta_{ij}}{{r}_{ij}},
  \label{Eq:dsalpha}
\end{equation}
where $\alpha$ and $\beta$ can take value 0 and 1 which correspond to x and y components, respectively. Now if particle-$i$ changes its position the angular displacement remains unaffected, i.e.~$\frac{\mathrm d \theta_i}{\mathrm d r^\alpha_i}=0$. Thus, the change in tangential displacement along $\beta$ due to the change in position of particle-$i$ along $\alpha$ only contributes in translations, and it can be written as (using~(\ref{Eq:dsalpha}))
\begin{equation}
  \frac{\mathrm d t^\beta_{ij}}{\mathrm d r^\alpha_i} = \Delta_{\alpha\beta} - \frac{r^\alpha_{ij} r^\beta_{ij}}{{r}^2_{ij}},
  \label{Eq:dsbdra}
\end{equation}
where $\Delta_{\alpha\beta}$ is the Kronecker delta which is one when $\alpha=\beta$, or else zero. Similarly, a change in rotational coordinates does not modify the particles relative distance, i.e.~$\frac{\mathrm d r^\beta_{ij}}{\mathrm d \theta_i}=0$. Thus, the change in tangential displacement along $\beta$ due to the change in $\theta_i$ is (from~(\ref{Eq:dsalpha}))
\begin{equation}
  \frac{\mathrm d t^\beta_{ij}}{\mathrm d \theta_i} = (-1)^\beta \sigma_i \frac{r^\alpha_{ij}}{{r}_{ij}}.
  \label{Eq:dsbdtheta}
\end{equation}
In the above equation $\alpha$ and $\beta$ are always different. Now the magnitude of tangential distance $t_{ij}$ can be obtained from the relation $t^2_{ij} = \sum_{\alpha} {t^\alpha_{ij}}^2$. Its differential follows $\mathrm d t_{ij} =  \sum_{\alpha} \frac{t^\alpha_{ij}}{t_{ij}} \mathrm d t^\alpha_{ij}$. The derivatives of tangential distance $t_{ij}$ with respect to $r^\alpha_{i}$ and $\theta_i$ can be expressed as
\begin{eqnarray}
  \label{Eq:dsmagdra}
  \frac{\mathrm d t_{ij}}{\mathrm d r^\alpha_i} &=&  \left(\frac{t^x_{ij}}{t_{ij}}\right)\frac{\mathrm d t^x_{ij}}{\mathrm d r^\alpha_i} + \left(\frac{t^y_{ij}}{t_{ij}}\right)\frac{\mathrm d t^y_{ij}}{\mathrm d r^\alpha_i}, \\
  \frac{\mathrm d t_{ij}}{\mathrm d \theta_i} &=&  \left(\frac{t^x_{ij}}{t_{ij}}\right)\frac{\mathrm d t^x_{ij}}{\mathrm d \theta_i} + \left(\frac{t^y_{ij}}{t_{ij}}\right)\frac{\mathrm d t^y_{ij}}{\mathrm d \theta_i}.
  \label{Eq:dsmagdtheta}
\end{eqnarray}
With the help of equations~(\ref{Eq:dsbdra}) and~(\ref{Eq:dsbdtheta}) we can solve the above two differential equations. As the tangential threshold is a linear function of overlap distance $\delta_{ij}$ (see~(\ref{Eq:s*})), it also gets modified due to a change in $r^\alpha_i$ as
\begin{equation}
  \frac{\mathrm d t^*_{ij}}{\mathrm d r^\alpha_i} = - \mu\left(\frac{k_n}{k_t}\right)\frac{r^\alpha_{ij}}{r_{ij}},
  \label{Eq:ds*dra}
\end{equation}
and it is unaffected by the change in rotation, i.e.~$\frac{\mathrm d t^*_{ij}}{\mathrm d \theta_i}=0$.
%%%%%%%%%%%%%%%%%%%%%%%%%%%%%%%%%%%%%%%%%%%%%%%%%%%%%%%%%%%%%%%%%%%%%%%%%%%%%%%%%%%

%%%%%%%%%%%%%%%%%%%%%%%%%%%%%%%%%%%%%%%%%%%%%%%%%%%%%%%%%%%%%%%%%%%%%%%%%%%%%%%%%%%
\subsection{Evaluation of ${\B J}$\label{sec:J}}
The derivative of tangential force (equation~(\ref{Eq:Fs})) with respect to $r^\alpha_{i}$:
\begin{eqnarray}
 &&\frac{\partial {F^{(t)}_{ij}}^\beta}{\partial r^\alpha_i} = -k_t\frac{\partial}{\partial r^\alpha_i}\left[\delta^{1/2}_{ij}\left(t^\beta_{ij} + \tilde{t} t^\beta_{ij} - {\tilde t}^2t^\beta_{ij} \right)\right]
  \nonumber \\
&=& -\frac{1}{2}\delta^{-1}_{ij}\frac{r^\alpha_{ij}}{r_{ij}}{F^{(t)}_{ij}}^\beta - k_t\delta^{1/2}_{ij} \times \nonumber \\
&&\left[ (1+{\tilde t}-{\tilde t}^2)\frac{\partial t^\beta_{ij}}{\partial r^\alpha_i} + ({\tilde t}^\beta - 2{\tilde t}{\tilde t}^\beta)\frac{\partial t_{ij}}{\partial r^\alpha_i} + (-{\tilde t}{\tilde t}^\beta+2{\tilde t}^2{\tilde t}^\beta)\frac{\partial t^*_{ij}}{\partial r^\alpha_i} \right] \nonumber \\
&&
  \label{Eq:dFsbdra}
\end{eqnarray}
Here we use the notation $\tilde t$ to represent the ratio $t_{ij}/t^*_{ij}$, and the notation ${\tilde t}^\beta$ for ${t_{ij}}^\beta/t^*_{ij}$. The expressions for all the three partial differentiation in~(\ref{Eq:dFsbdra}) are already shown in~(\ref{Eq:dsbdtheta}), (\ref{Eq:dsmagdra}), and (\ref{Eq:ds*dra}).

Similarly, the derivative of tangential force with respect to $\theta_{i}$ (using the same notation as above) can be found as
\begin{equation}
  \frac{\partial {F^{(t)}_{ij}}^\beta}{\partial \theta_i} = -k_t\delta^{1/2}_{ij} \left[ (1+{\tilde t}-{\tilde t}^2)\frac{\partial t^\beta_{ij}}{\partial \theta_i} + ({\tilde t}^\beta - 2{\tilde t}{\tilde t}^\beta)\frac{\partial t_{ij}}{\partial \theta_i} \right]
  \label{Eq:dFsbdtheta}
\end{equation}
From the above two equations it is then understood that if ${\B r}_{ij}$ and ${\B t}_{ij}$ are known the differential equations can be solved easily. When ${\tilde t}^\beta$ is negligible for all $\beta$, then ${\tilde t}\approx 0$. This translates to $\frac{\partial {F^{(t)}_{ij}}^\beta}{\partial \theta_i} = -(-1)^\beta k_t\sigma_i\delta^{1/2}_{ij}\frac{r_{ij}^\alpha}{r_{ij}}$ with $\alpha \neq \beta$, implying that even in the case of zero tangential displacement and therefore, zero tangential force, the above derivative can be finite.

%\subsection{Derivative of normal force}
The derivative of normal force (equation~(\ref{Eq:Fn})) with respect to $r^\alpha_{i}$:
\begin{eqnarray}
&&\frac{\partial {F^{(n)}_{ij}}^\beta}{\partial r^\alpha_i} = k_n \frac{\partial }{\partial r^\alpha_i} \left[ \delta^{3/2}_{ij} \frac{r^\beta_{ij}}{r_{ij}} \right]
  \nonumber \\
&&= k_n \delta^{1/2}_{ij}\left[ \Delta_{\alpha\beta}\frac{\delta_{ij}}{r_{ij}} - \frac{3}{2}\frac{r^\alpha_{ij} r^\beta_{ij}}{r^2_{ij}} -\left(\frac{\delta_{ij}}{r_{ij}}\right) \frac{r^\alpha_{ij} r^\beta_{ij}}{r^2_{ij}} \right],
  \label{Eq:dFnbdra}
\end{eqnarray}
where $\Delta_{\alpha\beta}$ is the Kronecker delta. The derivative of total force which reads:
\begin{eqnarray}
\label{Eq:dFbdra}
\frac{\partial {F_{ij}}^\beta}{\partial r^\alpha_i} &=& \frac{\partial {F^{(n)}_{ij}}^\beta}{\partial r^\alpha_i} + \frac{\partial {F^{(t)}_{ij}}^\beta}{\partial r^\alpha_i} \\
\frac{\partial {F_{ij}}^\beta}{\partial \theta_i}  &=& \frac{\partial {F^{(t)}_{ij}}^\beta}{\partial \theta_i}
\label{Eq:dFbdtheta}
\end{eqnarray}
can be solved using~(\ref{Eq:dFnbdra}),~(\ref{Eq:dFsbdra}), and~(\ref{Eq:dFsbdtheta}).
%%%%%%%%%%%%%%%%%%%%%%%%%%%%%%%%%%%%%%%%%%%%%%%%%%%%%%%%%%%%%%%%%%%%%%%%%%%%%%%%%%%

%%%%%%%%%%%%%%%%%%%%%%%%%%%%%%%%%%%%%%%%%%%%%%%%%%%%%%%%%%%%%%%%%%%%%%%%%%%%%%%%%%%
%subsection{Derivative of Torque}
The torque of particle-$j$ due to tangential force ${\B F^{(t)}}_{ij}$ is ${\B T}_j = -\sigma_j\left({\hat r}_{ij} \times {\B F^{(t)}}_{ij}\right) \equiv \sigma_j \tilde{\B{T}}_{ij}$.
In 2D, $\tilde{\B T}_{ij}$ has only z-component:
\begin{equation}
  {\tilde T}_{ij}^z = - \left[ \left(\frac{x_{ij}}{r_{ij}}\right){F^{(t)}_{ij}}^y - \left(\frac{y_{ij}}{r_{ij}}\right){F^{(t)}_{ij}}^x\right].
  \label{Eq:Tz}
\end{equation}
The derivative of ${\tilde T}_{ij}^z$ then becomes:
\begin{equation}
\begin{split}
  \label{Eq:dTdra}
  \frac{\partial {\tilde T}_{ij}^z}{\partial r^\alpha_i} =& -\left(\frac{\delta_{\alpha x}}{r_{ij}} - \frac{x_{ij}r_{ij}^\alpha}{r_{ij}^3}\right){F^{(t)}_{ij}}^y -  \left(\frac{x_{ij}}{r_{ij}}\right)\frac{\partial {F^{(t)}_{ij}}^y}{\partial r^\alpha_i} + \\
&\left(\frac{\delta_{\alpha y}}{r_{ij}} - \frac{y_{ij}r_{ij}^\alpha}{r_{ij}^3}\right){F^{(t)}_{ij}}^x +  \left(\frac{y_{ij}}{r_{ij}}\right)\frac{\partial {F^{(t)}_{ij}}^x}{\partial r^\alpha_i},
\end{split}
\end{equation}
where $\delta_{\alpha x}$ (similarly,  $\delta_{\alpha y}$) is the Kronecker delta, such that $\delta_{x x}=1$ and $\delta_{y x}=0$, and
\begin{equation}
  \frac{\partial {\tilde T}_{ij}^z}{\partial \theta_i} =  - \left[ \left(\frac{x_{ij}}{r_{ij}}\right)\frac{\partial {F^{(t)}_{ij}}^y}{\partial \theta_i} - \left(\frac{y_{ij}}{r_{ij}}\right)\frac{\partial {F^{(t)}_{ij}}^x}{\partial \theta_i}\right]
  \label{Eq:dTdtheta}
\end{equation}
The above two differential equations can be solved using~(\ref{Eq:dFsbdra}), and~(\ref{Eq:dFsbdtheta}). If the tangential displacement $t_{ij}^\beta$ is negligible compared to the threshold $t_{ij}^*$, i.e., ${\tilde t}^\beta \approx 0$ for all $\beta$. This results in ${\tilde t}\approx 0$. Therefore, $\frac{\partial {\tilde T}_{ij}^z}{\partial \theta_i} = k_t\sigma_i\delta_{ij}^{1/2}$.
%%%%%%%%%%%%%%%%%%%%%%%%%%%%%%%%%%%%%%%%%%%%%%%%%%%%%%%%%%%%%%%%%%%%%%%%%%%%%%%%%%%

%%%%%%%%%%%%%%%%%%%%%%%%%%%%%%%%%%%%%%%%%%%%%%%%%%%%%%%%%%%%%%%%%%%%%%%%%%%%%%%%%%%
\subsection{Jacobian\label{sec:Jcomponents}}
The dimension of Jacobian operator $J$ is force over length. To be consistent with the dimension we redefine the torque $T$ and rotational coordinate $\theta$ as
\begin{equation}
  {\tilde T}_i = \frac{T_i}{\sigma_i}, \ \ \ \textrm{and} \ \ \ {\tilde \theta}_i = \sigma_i\theta_i
  \label{Eq:rescaledT}
\end{equation}
In addition, the dynamic matrix has a contribution from the moment of inertia $I_i=I_0m_i\sigma_i^2$ as $\Delta{\B \omega}_i={\B T}_i/I_i\Delta t$. In our calculation, we assume that mass $m_i$ and $I_0$ both are one. The remaining contribution of $I_i$, i.e. $\sigma_i^2$, is taken care of by rescaling the torque and the angular displacement as ${\tilde T}_i$ and ${\tilde \theta_i}$~(\ref{Eq:rescaledT}). For $I_0\neq 1$, the contribution of $I_0$ can be correctly anticipated if we rewrite~(\ref{Eq:dsdt}) as below:
\begin{equation}
  \frac{\mathrm d {\B t}_{ij}}{\mathrm d t} = {\B v}_{ij} - {\B v}^n_{ij} + \frac{1}{I_0}\hat{r}_{ij}\times(\sigma_i{\B \omega}_i + \sigma_j{\B \omega}_j),
  \label{Eq:dsdtnew}
\end{equation}

$J$ essentially contains {\bf four} different derivatives:
\begin{itemize}
\item First type: Derivative of force with respect to the position of particles:
  \begin{equation}
    \label{Eq:Jabij}
    \begin{split}
      & J^{\alpha\beta}_{ij} = \sum_{k=0; k\neq j}^{N-1}\frac{\partial F^\beta_{kj}}{\partial r^\alpha_{i}} = \frac{\partial F^\beta_{ij}}{\partial r^\alpha_{i}}, \ \ \ \textrm{for}\ \ i\neq j \\
      & J^{\alpha\beta}_{ii} = \sum_{j=0; j\neq i}^{N-1}\frac{\partial F^\beta_{ji}}{\partial r^\alpha_{i}} = -\sum_{j=0; j\neq i}^{N-1} J^{\alpha\beta}_{ij},
    \end{split}
  \end{equation}
where $N$ is the total number of particles. $J^{\alpha\beta}_{ij}$ is symmetric if we change pairs, i.e.: $J^{\alpha\beta}_{ij}=J^{\alpha\beta}_{ji}$, however the symmetry is not guaranteed with the interchange of $\alpha$ and $\beta$.
\item Second type: Derivative of force with respect to rotational coordinate:
   \begin{equation}
    \label{Eq:Jbij}
    \begin{split}
      & J^{\beta}_{ij} = -\sum_{k=0; k\neq j}^{N-1}\frac{\partial F^\beta_{kj}}{\partial \tilde\theta_{i}} = -\frac{\partial F^\beta_{ij}}{\partial \tilde\theta_{i}}, \ \ \ \textrm{for}\ \ i\neq j \\
      & J^{\beta}_{ii} = -\sum_{j=0; j\neq i}^{N-1}\frac{\partial F^\beta_{ji}}{\partial \tilde\theta_{i}} = \sum_{j=0; j\neq i}^{N-1} J^{\beta}_{ij}.
    \end{split}
   \end{equation}
   The negative sign makes sure that in stable systems all the eigenvalues are positive. $J^{\beta}_{ij}$ is asymmetric: $J^{\beta}_{ij}=-J^{\beta}_{ji}$.
 \item Third type: Derivative of torque with respect to position:
   \begin{equation}
     \label{Eq:Jaij}
     \begin{split}
       & J^{\alpha}_{ij} = \sum_{k=0; k\neq j}^{N-1}\frac{\partial {\tilde T}_{kj}^z}{\partial r^\alpha_{i}} = \frac{\partial {\tilde T}_j}{\partial r^\alpha_{i}}, \ \ \ \textrm{for}\ \ i\neq j \\
       & J^{\alpha}_{ii} = \sum_{j=0; j\neq i}^{N-1}\frac{\partial {\tilde T}_{ji}^z}{\partial r^\alpha_{i}} = \sum_{j=0; j\neq i}^{N-1} J^{\alpha}_{ij}.
     \end{split}
   \end{equation}
$J^{\alpha}_{ij}$ is also asymmetric: $J^{\alpha}_{ij}=-J^{\alpha}_{ji}$.
 \item Fourth type: Derivative of torque with respect to rotational coordinate:
   \begin{equation}
     \label{Eq:Jij}
     \begin{split}
       & J_{ij} = -\sum_{k=0; k\neq j}^{N-1}\frac{\partial {\tilde T}_{kj}^z}{\partial \tilde\theta_{i}} = -\frac{\partial {\tilde T}_j}{\partial \tilde\theta_i}, \ \ \ \textrm{for}\ \ i\neq j \\
       & J_{ii} = -\sum_{j=0; j\neq i}^{N-1}\frac{\partial {\tilde T}_{ji}^z}{\partial \tilde\theta_{i}} = -\sum_{j=0; j\neq i}^{N-1} J_{ij}.
     \end{split}
   \end{equation}
   The negative sign makes sure that in stable systems all the eigenvalues are positive. $J_{ij}$ is symmetric: $J_{ij}=J_{ji}$.
\end{itemize}
%%%%%%%%%%%%%%%%%%%%%%%%%%%%%%%%%%%%%%%%%%%%%%%%%%%%%%%%%%%%%%%%%%%%%%%%%%%%%%%%%%%

%%%%%%%%%%%%%%%%%%%%%%%%%%%%%%%%%%%%%%%%%%%%%%%%%%%%%%%%%%%%%%%%%%%%%%%%%%%%%%%%%%%
\subsection{Arrangement of Jacobian matrix\label{sec:Jarrangement}}
In two dimension $D=2$, for $N$ particles the total number of elements in $J$ is $(D+1)N\times(D+1)N$. In the matrix, first $DN\times DN$ elements contain the first type of force derivative, i.e. $J^{\alpha\beta}_{ij}$. Here the row-index $ro$ and column-index $co$ of $J$ runs in the range $0 \leq ro < DN$ and $0 \leq co < DN$. Rows from $DN \leq ro < (D+1)N$ and columns $0 \leq co < DN$ of $J$ contain $J^\beta_{ij}$, i.e., the second type of derivative. Rows from $0 \leq ro < DN$ and columns $DN \leq co < (D+1)N$ of $J$ contain the third type $J^\alpha_{ij}$. Finally, rows from $DN \leq ro < (D+1)N$ and columns $DN \leq co < (D+1)N$ of $J$ hold $J_{ij}$, i.e., the fourth type of derivative. For a fixed type of derivative, at a fixed row, the column-index first runs over $j$ starting from 0 to $N-1$. Then $\beta$ is incremented, if it exists for that particular derivative type. Similarly, at a fixed column, row-index first runs over $i \in [0, N)$ and then $\alpha$ is incremented.

\bibliography{ALL}

\end{document}